\documentstyle[12pt,aasms4]{article}
\begin{document}

\title{A Slower Superluminal Velocity for the Quasar 1156+295}

\author{B. Glenn Piner\altaffilmark{1,2} and Kerry A. Kingham} 

\affil{U.S. Naval Observatory, Earth Orientation Dept., 3450 Massachusetts
Ave,
Washington D.C. 20392}

\altaffiltext{1}{Department of Astronomy, University of Maryland,
College Park MD 20742}
\altaffiltext{2}{NASA/Goddard Space Flight Center, Code 661, Greenbelt, MD
20771
}

\begin{abstract}

As part of an ongoing effort to observe high energy $\gamma$-ray blazars with VLBI,
we have produced 8 and 2 GHz VLBI images, at ten epochs spanning the
years 1988 to 1996, of the quasar 1156+295.
The VLBI data have
been taken from the Washington VLBI correlator's geodetic database.  
We have detected four components and have measured their apparent speeds to be
8.8 $\pm$ 2.3, 5.3 $\pm$ 1.1, 5.5 $\pm$ 0.9, and 3.5 $\pm$ 1.2 $h^{-1}c$ from the outermost
component inwards. ($H_{0}=100h$ km s$^{-1}$ Mpc$^{-1}$, $q_{0}$=0.5 throughout paper).
These velocities contradict a previously published very high superluminal velocity
of 26 $h^{-1}c$ for this source.

\end{abstract}
\newpage

\section{Introduction}

The source 1156+295 is a high polarization quasar (Wills et al. 1992), and
an optically violent variable quasar with a redshift of z=0.73 (Hewitt \& Burbidge 1989). 
Tornikoski et al. (1994) discuss the monitoring of this source
at optical wavelengths at the Rosemary Hill Observatory,
and at the higher radio frequencies by the Mets\"{a}hovi group.
It has also been monitored at lower radio frequencies by the Michigan group
(Aller et al. 1985).  This quasar is an emitter of high energy $\gamma$-rays; 
EGRET detections of this source are reported by Thompson et al. (1995).

Previous VLBI observations of this source are presented by McHardy et al. (1993, 1990),
who measured an apparent velocity of 26 $h^{-1}c$
based on four images at three different frequencies from 1986 to 1988.
This velocity made 1156+295 by far the fastest known superluminal source.
Vermeulen \& Cohen (1994) remark that at the velocity reported by 
McHardy et al. (1993, 1990) 1156+295
does not even appear to be the tail end of a continuous distribution; its speed is 2.5 times
higher than that of the next fastest object in its redshift bin on their $\mu$-z diagram.
Since the apparent velocity can be used to set a lower limit on the Lorentz factor, 
$\Gamma \geq (\beta_{app}^{2} +1)^{1/2}$, this apparent velocity also implies a very
high Lorentz factor for the superluminal component of $\Gamma \geq 26$.

\section{VLBI Observations}
  
The VLBI results presented here were obtained as part of a project to
study EGRET $\gamma$-ray blazars using archived geodetic Mark III VLBI
observations processed at the Washington VLBI Correlator Facility located
at the U.S. Naval Observatory (USNO).  Details of the geodetic VLBI observations
and the data reduction and model fitting procedures are discussed by
Piner \& Kingham (1997).  We have imaged 1156+295 at 8 and 2 GHz at ten
different epochs: 1988 November 25, 1989 April 11, 1989 September 26, 1989 December 18,
1990 June 29, 1992 June 30, 1993 May 7, 1993 July 14, 1996 January 9, and 1996 March 30.
Up to three geodetic VLBI experiments were combined to produce
single images, with a total of sixteen experiments being used to produce images at
ten epochs.  When experiments were combined the maximum time separation between
experiments was limited to be less than 45 days, so that the time span
covered by the combined experiments would be less than the time scale for
intrinsic source structure changes.
A total of 25 different antennas worldwide 
were used among all of the imaged observations, with a maximum of six being used
for a single experiment.

\section{Motion in 1156+295}
Figure 1 shows representative 8 and 2 GHz images from the beginning, middle,
and end of the time span covered by the observations.  The other 
fourteen images are not presented due to space considerations.
Note the differences in scale and resolution between the 8 and 2 GHz images.
The outer components are more easily seen in the 2 GHz images and are not
as easily seen in the 8 GHz images due to these image's high resolution and the
dearth of short baselines in these observations.  The spectrum of the components
also steepens as the components move out, so that as time goes on they become
less easily detected in the 8 GHz images.
The earliest 2 GHz image, from 1988 November 25, is from about the same time as the latest image
presented by McHardy et al. (1990), and agrees well with their Figure 6$a$.
Two components, labeled C1 and C2, are visible in Figure 1$a$ in addition to
the bright core component.  The outermost component, C1, is only detected in
those 2 GHz observations with the best dynamic range and $(u,v)$ plane coverage, and is, 
for example, not seen in the 1992 June 30 2 GHz image.  Component C2 is the brightest
component at 2 GHz and is seen in all ten 2 GHz images and two 8 GHz images.
This component corresponds to the component that McHardy et al. (1990)
measured at 5 GHz on 1988 November 12.
The next innermost component, C3, is initially visible only in the 8 GHz images,
but later becomes detectable in the 2 GHz images as it moves outward.  The innermost
component, C4, is detectable only in the later 8 GHz images.

Figure 2 shows the motions of the components over time.  
Component positions measured at 2 GHz are plotted as squares and their motions
are fitted with solid lines, while component positions measured at 8 GHz are plotted
as diamonds and their motions are fitted with dotted lines.
The component positions were obtained by fitting Gaussian model components
to the observed visibilities, and
the errors in
the component positions are taken to be one quarter of the maximum projection
of the beam FWHM along the
direction from the core to the component's center.  
In addition to observations at the ten epochs mentioned above, we have also plotted
component positions at 1994 July 8 given by Fey, Clegg, \& Fomalont (1996) 
from a VLBA observation at 8 and 2 GHz.
We find that measurements of component positions at 8 GHz are on average displaced
outwards from the same component position measured at 2 GHz by 0.65 milliarcseconds (mas).
This frequency-dependent separation has been seen by other authors 
(e.g. Biretta, Moore, \& Cohen 1986), 
and is possibly
due to gradients in magnetic field and electron density, such that
the $\tau =1$ surface moves progressively inward at higher frequencies.
If we assume that the components move with constant velocity, then
performing a least-squares fit of component positions to a straight line gives the apparent
velocities of separation.  The velocities obtained are 
8.8 $\pm$ 2.3, 5.3 $\pm$ 1.1, 5.5 $\pm$ 0.9, and 3.5 $\pm$ 1.2 $h^{-1}c$ for C1 to C4
respectively, where the average of the 8 and 2 GHz velocities has been used for C2 and C3.
The one sigma ranges for the ejection times of the two most recent components are
1984 July to 1986 July for C3, and 1989 June to 1992 January for C4.
The velocities and standard errors are consistent with each of the components moving
with similar velocity; $\sim 5.2 h^{-1}c$.
This lowers the required Lorentz factor for the superluminal components from
$\Gamma \geq 26$ to $\Gamma \geq 5$.

The inner components are not moving significantly faster than the outer components, which
leads us to believe that the discrepancy between the velocities measured by us and that
measured by McHardy et al. (1993, 1990) is probably not due to a sudden deceleration of
the component.
We speculate that the anomalously high velocity measured by McHardy et al. (1993, 1990)
may be due to a combination of several factors.  Their earliest observed point is at
their highest observing frequency, 22 GHz, and the position they measure is consistent
with the position we estimate for C3 at that time, taking frequency-dependent separation
into account.  Since their latest point at 5 GHz is consistent with our measured 
position for the component C2, we speculate that they may have identified these two
different components as the same component.  This is a danger when connecting components
from images made at different frequencies.  Their central two points fall between the
positions we estimate for C3 and C2 at those times.  This may be partly due to an
underestimation of the true error bars on the separation, and partly due to a shift of
the apparent core position northwards due to a merger of C3 and the core at the frequencies
of these observations, 5 and 11 GHz.

We conclude that 1156+295 does not show the extremely rapid superluminal expansion
previously measured for this source, and instead has an apparent superluminal velocity
much closer to the average for core dominated quasars.  This reduces the highest observed 
speed for superluminal sources from 26 $h^{-1}c$ to 21 $h^{-1}c$ for
component K2 of 1308+326 (Gabuzda et al. 1993).

\newpage
\begin{center}
\bf{Figure Captions} \\
\end{center}

\figcaption{($a$)-($c$): 2 GHz VLBI images of 1156+295. 
($d$)-($f$): 8 GHz VLBI images of 1156+295.
Relative right ascension and declination are plotted with tickmark spacings of 1 mas.
Contour levels are 2\%, 4\%, 8\%, 16\%, 32\%, and 64\% 
of the peak brightness.
Additional contour levels are $-$0.7\%, 0.7\%, and 1.2\%
for ($a$), ($d$), and ($f$);
$-$0.55\%, 0.55\%, and 1\% for ($b$);
and $-$0.5\%, 0.5\%, and 1\% for ($c$) and ($e$).
The FWHMs in mas and position angles of the restoring beams are
2.52 x 2.07 at $-63\arcdeg$, 3.36 x 2.05 at $-72\arcdeg$,
2.77 x 2.25 at $-6\arcdeg$, 0.70 x 0.54 at $-66\arcdeg$,
0.94 x 0.60 at $-73\arcdeg$, and 0.73 x 0.59 at $-10\arcdeg$ for ($a$)-($f$) respectively.
The peak brightness levels are
1.67, 1.37, 1.15, 1.27, 1.34, and 1.13 Jy beam$^{-1}$ respectively.
At least one negative contour is shown in each image.
The centers of the fitted component positions are marked with asterisks.} 

\figcaption{Motion of components in 1156+295.  The vertical 
axis shows the separation in mas of the center
of the component from the presumed core. 
Component positions measured at 2 GHz are plotted as squares and
positions measured at 8 GHz are plotted as diamonds.
The solid lines represent the best fit of the 2 GHz positions to motion
with constant velocity, and the dotted lines represent the same fits
for the 8 GHz positions.} 

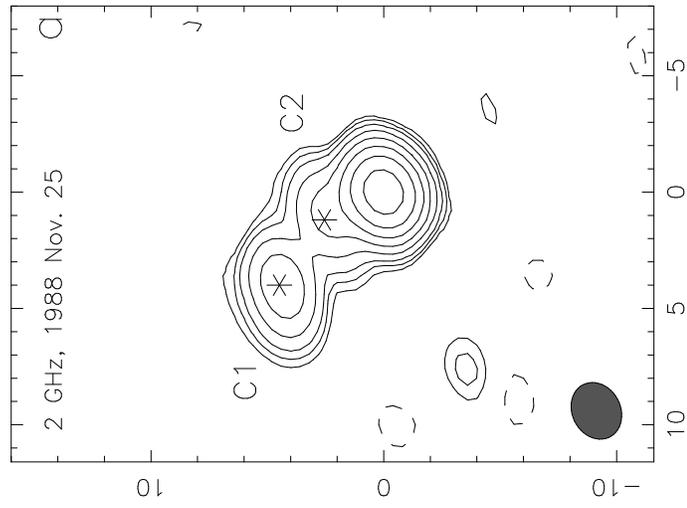
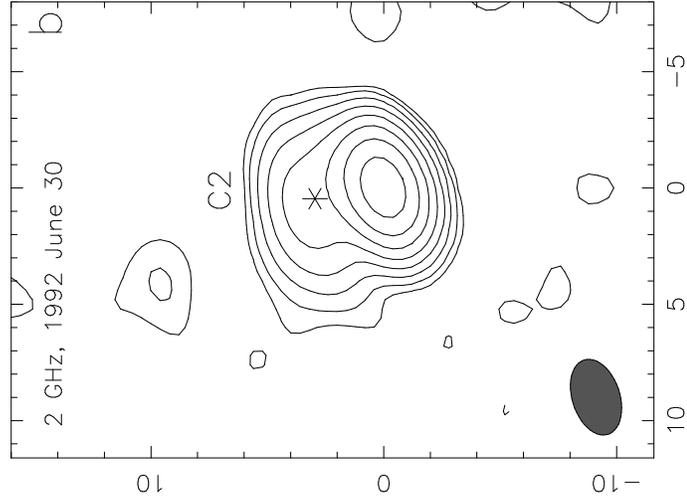
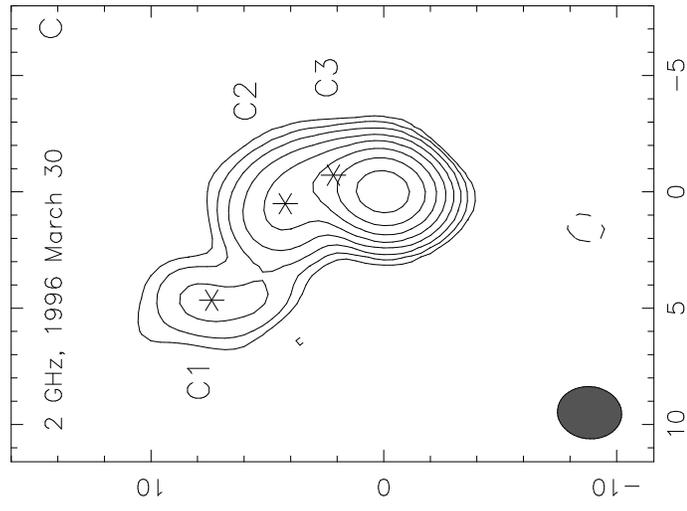
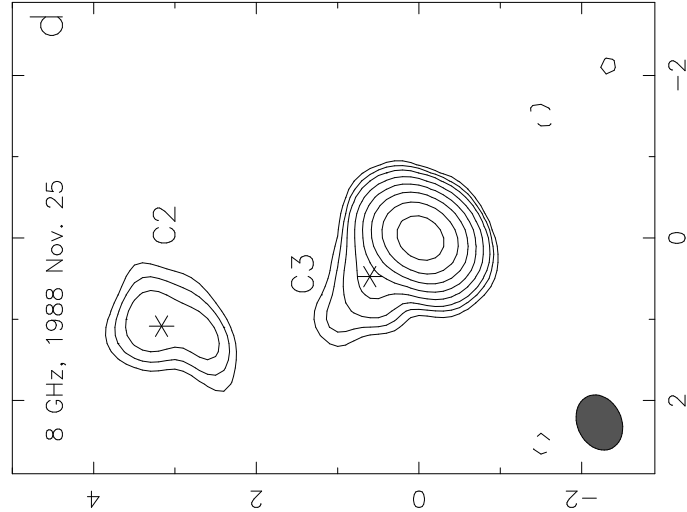
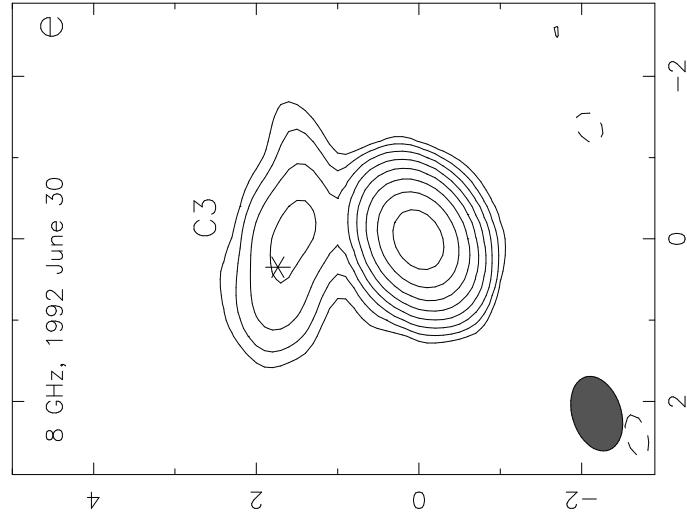
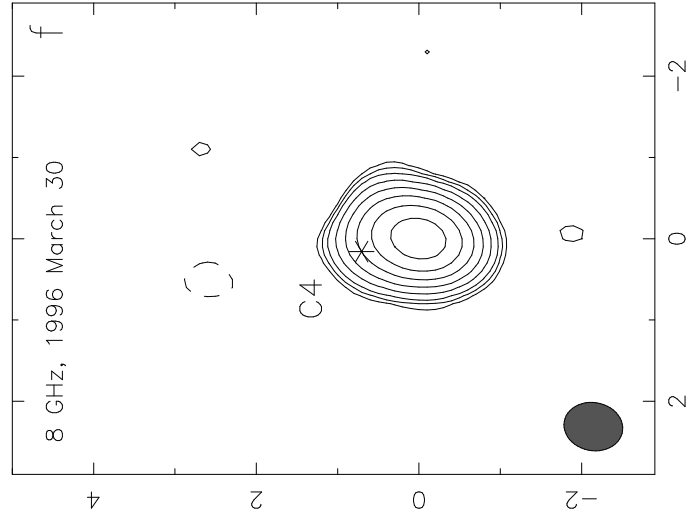
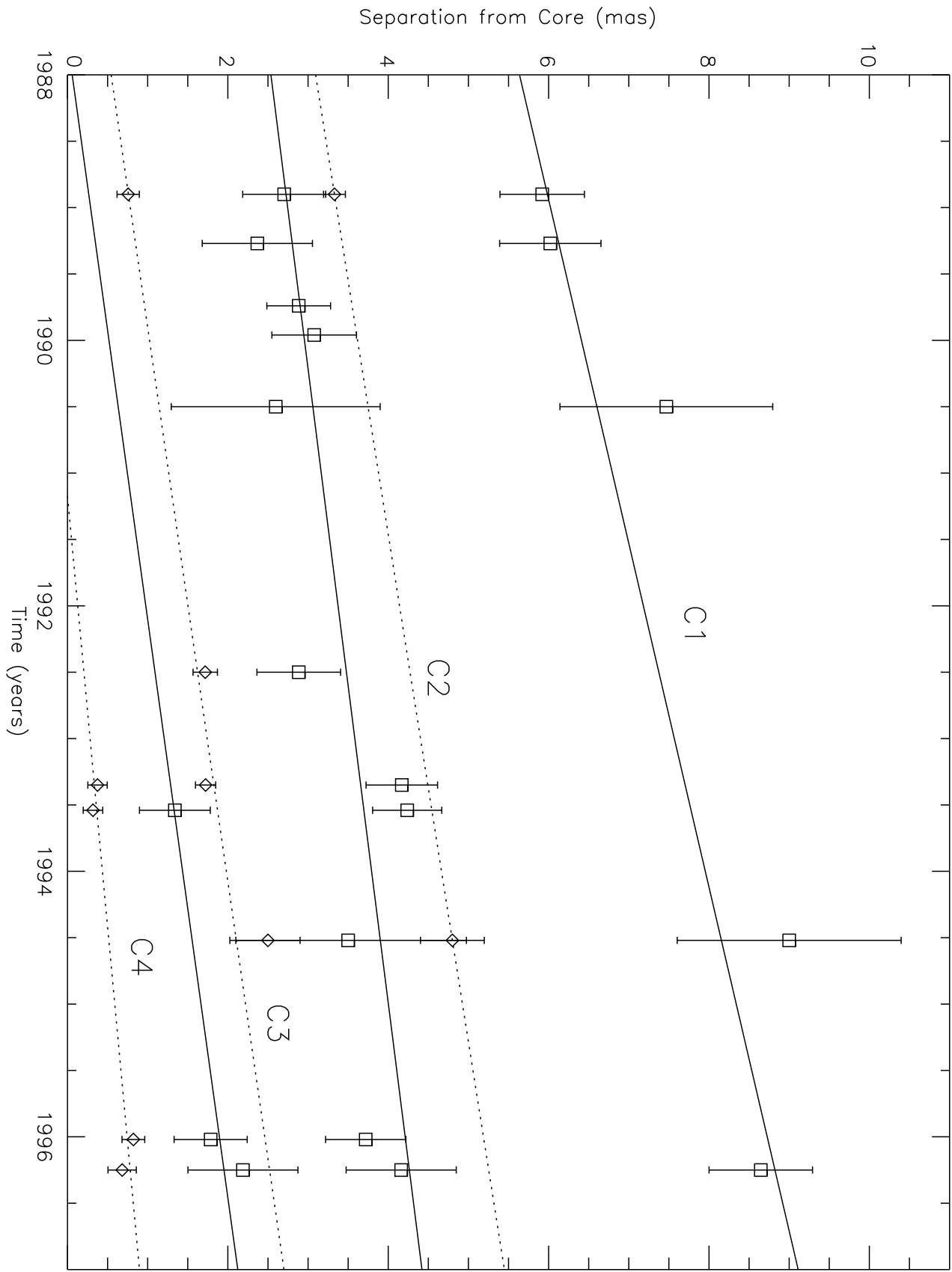
\end{document}